\documentclass[11pt]{article}

\usepackage[left=1in,top=1in,right=0.9in,bottom=0.9in,nohead,nofoot]{geometry}

 \usepackage{amsthm,amssymb}
 \usepackage{amsmath}
 \usepackage{algorithm}
\usepackage{algpseudocode}

\newtheorem{theorem}{Theorem}[section]
\newtheorem{lemma}[theorem]{Lemma}

\newtheorem{definition}[theorem]{Definition}

\newtheorem{remark}[theorem]{Remark}

\newcommand*\fprog{f_{prog}}

\newcommand{\todoI}[1]{{}}

\newcommand{\eps}{\varepsilon}
\newcommand{\beep}{\textsc{Beep }}
\newcommand{\degr}{\Delta}
\newcommand{\local}{\textsc{Local }}
\newcommand{\epsack}{\varepsilon_{ack}}

\makeatletter
\newcommand*{\StartNewContent}{%
    \let\OrigLabel\label%
    \let\OrigRef\ref%
    \renewcommand{\label}[1]{\OrigLabel{FULL:##1}}%
    \renewcommand{\ref}[1]{\OrigRef{FULL:##1}}%
    \renewcommand{\label@in@display}[1]{%
        \ifx\df@label\@empty\else
            \@amsmath@err{Multiple \string\label's:
                label '\df@label' will be lost}\@eha
        \fi
        \gdef\df@label{FULL:##1}%
    }%
}
\makeatother

\newif\iffull
\newif\ifshort

\begin{document}

\date{} 
\title{Beeping a Maximal Independent Set Fast}

\author{
  Stephan Holzer\footnote{Supported by: AFOSR Contract Number FA9550-13-1-0042, NSF Award 0939370-CCF, NSF Award CCF-1217506, and NSF Award CCF-AF-1461559.}\\
  \texttt{\small holzer@mit.edu}\\
	\small MIT
  \and
  Nancy Lynch${}^*$\\
  \texttt{\small lynch@csail.mit.edu}\\
	MIT
}

\maketitle

\begin{abstract}
We adapt a recent algorithm by Ghaffari~\cite{G16} for computing a Maximal 
Independent Set in the \local model, so that it works in the 
significantly weaker \beep model.  For networks with 
maximum degree $\Delta$, our algorithm terminates locally within time 
$O((\log \degr + \log (1/\eps)) \cdot \log(1/\eps))$, with probability at least $1 - \eps$.\footnote{Errata note: In our brief announcement~\cite{holzer2016brief}, we claimed as a side-effect of our local bound, the analysis of~\cite{G16} can be used to show that this algorithm terminates globally within time $O(\log^2 \degr) + 2^{O(\sqrt{\log \log n})}$ with high probability in $n$, the number of nodes in the network. While it is unknown whether this bound can be achieved, it is not clear that it can be derived via the graph-scattering technique used in~\cite{G16} in combination with the deterministic algorithm of~\cite{DBLP:conf/stoc/PanconesiS92}. At least these techniques cannot be translated in the desired time in the beeping model as we thought of. The main reason is that in~\cite{DBLP:conf/stoc/PanconesiS92} nodes exchange more information than the \beep model can handle in the time we hoped to achieve. Studying local complexity is of interest by itself as recently demonstrated by the papers cited in the abstract.}  The key idea of 
the modification is to replace explicit messages about transmission 
probabilities with estimates based on the number of received 
messages.

After the successful introduction (and implicit use) of local analysis, e.g., in~\cite{barenboim2016locality,chung2014distributed,G16,halldorsson2015local-podc}, we study this concept in the \beep model for the first time. 

By doing so, we improve over local bounds that are implicitly derived from previous work (that uses traditional global analysis) on computing a Maximal Independent Set in the \beep model for a large range of values of the parameter $\degr$. At the same time, we show that our algorithm in the \beep model only needs to pay a $\log(1/\eps)$ factor in the runtime compared to the best known MIS algorithm in the much more powerful \local model. We demonstrate that this overhead is negligible, as communication via beeps can be implemented using significantly less resources than communication in the \local model. In particular, when looking at implementing these models~\cite{DBLP:journals/adhoc/KhabbazianKKL14}, one round of the \local model needs at least $O(\Delta)$ time units, while one round in the \beep model needs $O(\log\Delta)$ time units, an improvement that diminishes the loss of a $\log(1/\eps)$ factor in most settings.  
\end{abstract}

\newpage 

\shortfalse
\fulltrue 

\section{Introduction}
Computing a Maximal Independent Set (MIS) is a widely studied problem in distributed computing theory. One of the weakest models of communication in which this problem has been studied is the \beep model, e.g.,~\cite{beep13,cornejo2010deploying,feedback}. In this model, nodes can only send a beep or listen in order to communicate, i.e., no sender-collision detection is available. In the version of this model with synchronized clocks, collision detection, and simultaneous wakeup, the authors of~\cite{beep13} showed how to compute an MIS w.h.p. in time $O(\log^2 n)$ by adapting a classical MIS algorithm by Luby~\cite{Luby86}\footnote{They also consider other settings and argue that the assumption that all nodes wake up at the same time can be removed while keeping the same runtime.}. Subsequently~\cite{feedback} showed how to improve this bound to $O(\log n)$ by adapting an improved version of Luby's algorithm and tuning the probabilities of nodes joining the MIS over time. 

Local analysis is a concept that was recently introduced or implicitly used, e.g., in~\cite{barenboim2016locality,chung2014distributed,G16,halldorsson2015local-podc}. While traditional global complexity
guarantees, that with high probability all nodes terminate after a certain runtime (that typically depends on the size $n$ of the network), local complexity guarantees that each particular node $v$ terminates with probability $1-\eps$ after a number of rounds, that typically depends only on the degree of the node (or network) and the probability $1-\eps$ that needs to be achieved. Local complexity is of interest in settings, where nodes could immediately continue with other computations once they found their own part of the solution to a problem. Constant probabilities often suffice, which results in speedups. This was demonstrated, e.g., in~\cite{halldorsson2015local-podc} that uses iterated computations of MIS in the SINR model, where only a very low local success probability is required that ultimately yielded an improvement of global broadcast in the SINR model, as previous approaches always worked with global w.h.p. guarantees. Further motivation on local analysis as an addition to global analysis can be found in~\cite{G16}.

We study the concept of local analysis in the \beep model for the first time. This analysis improves over the canonical local complexity that can be derived from state-of-the-art algorithm and analysis~\cite{feedback} for a large range of values of the parameter $\Delta$ by reducing the runtime from $O(\log n)$ to $O((\log \degr + \log (1/\eps)) \cdot \log(1/\eps))$ as we focus on local termination and local correctness:
\begin{definition}[Local correctness of MIS algorithms]
Any node $v$ can produce output \textit{IN} or \textit{OUT} indicating $v$'s membership to the MIS. This output cannot be revised once it is made. If a node $v$ outputs \textit{IN}, then at that moment none of its neighbors are IN and none of its neighbors will output IN later. If a node outputs OUT, then at that moment some neighbor is already IN the MIS.
\end{definition}

\begin{theorem}[Local correctness (safety property)] \label{thm:local-correctness} 
In our Algorithm of Section~\ref{sec:beepalgo}, when a node $v$ terminates, it has made its (locally correct) decision whether it is in the MIS or not. 
\end{theorem}
Note that this safety property is conditioned on the termination of the algorithm and the next Theorem states that the algorithm terminates w.h.p..

\begin{theorem}[Local termination complexity (liveness property)] \label{thm:local-restate2}
In our Algorithm of Section~\ref{sec:beepalgo}, for each node $v$, the probability that node $v$ terminates within the first $O((\log \degr + \log (1/\eps)) \cdot \log(1/\eps))$ slots and makes a locally correct decision is at least $1-\eps$. This holds even if the outcome of the coin tosses outside $N^{+}_{2}(v):=V\setminus N_1(v)$ are determined adversarially.
\end{theorem}
We obtain this bound by adapting Ghaffari's algorithm~\cite{G16} for the \local model to work in the \beep model. The key idea in the proof and algorithm of Theorem~\ref{thm:local-restate2} is 
that, instead of maintaining full information about its 
neighbors' states, a node keeps a single binary 
estimate for the aggregate state of its entire neighborhood. In particular, the agents in the 
\beep algorithm in this paper estimate probabilities by observing their 
neighbors' probabilistically-generated transmission behavior. This results in improved runtimes and we show that the performance of the algorithms is close to the case in which the agents have exact information. If we shift our focus away from the pure notation of communication rounds or slots, it turns out that in many cases our new algorithms in the \beep model are even more efficient than the original \local model algorithm. The intuition for this is that
it takes $\Omega(B\Delta)$ time units to emulate one round of the \local model in the \beep model in case the message size is $B$, whereas each time slot of the \beep model can be emulated by one round in the \local model. 

Our approach is partly inspired by recent research on biological distributed algorithms such as~\cite{musco2016ant}, social networks, and other new forms of distributed algorithms.  In such systems, agents do not obtain precise information about other agents'  preferences and tendencies, and estimate these from observing their behavior via sampling. It is interesting to understand how to improve efficiency by using stochastic information that arises from sampling distributions, rather than collecting exact information on these distributions. 

\begin{remark}
Note that this local bound is only a factor of $O(\log(1/\eps))$ larger than the state-of-the-art $O(\log \Delta + \log (1/\eps))$ bound in the \local model~\cite{G16}. 
\end{remark}

\subsection{Motivation, Related Work and Our Contribution}
As pointed out above, communication in the \local model and \beep networks and their true implementation cost differ significantly. This requires us to be very careful when translating \local algorithms into \beep networks in order to not lose the strength of the techniques behind them and to obtain algorithms that are in many settings even more efficient due to the simpler nature of the \beep model. The key difference between the models is, that within one communication round in the \local model, a node can exchange arbitrarily large and different messages with all of its neighbors at the same time, while in \beep networks a node can beep or not beep (this conveys less information than a conventional $1$-bit message, where one has the option to send $0/1$, or not send at all) and can only send or receive one message in the same slot. In the \beep model, this message is restricted to contain one beep of information and therefore each round in the \beep model needs much less resources than a round in the \local model. 

A straightforward and unfortunately inefficient way to emulate the behavior of one round in the \local model in the \beep model by performing a \emph{local broadcast} of each node's message that it would send in the \local model. This takes $\Theta(B\Delta + \Delta \cdot poly\log (\Delta/\epsack))$ time slots, where $B$ denotes the size of the message in the \local model, and  the local broadcast is guaranteed to succeed with probability $1-\epsack$ (ack stands for acknowledged broadcast). 

However, this simple technique of translating algorithms from the \local model into the \beep model is highly inefficient for algorithms that have small \local runtimes, as $\Delta$ is a factor in the \beep runtime. In particular, this is much higher than the (local termination) complexity of $O((\log \degr + \log (1/\eps)) \cdot \log(1/\eps))$ that we achieve (note that $\eps$ is often a constant when using local complexity), which can be exponentially faster than the factor $\Delta$ lost by applying the simple transfer technique described above. 

Finally, we remark that readers familiar with the $O(\log^*n)$ MIS algorithm of Schneider and Wattenhofer~\cite{DBLP:conf/podc/SchneiderW08} in the \local model for Bounded Growth Graphs might wonder why we did not translate their algorithm, as Bounded Growth Graphs capture most wireless network topologies in which the  \beep model is used. We show in Theorem~\ref{thm:local-BGG} that their algorithm cannot be translated to the \beep model without major modifications without paying a $\Delta$ factor in the runtime and would therefore be exponentially worse than our solution.

\begin{table}[h]
\begin{center}
\begin{tabular}{l|c|c|l}
\multicolumn{4}{l}{{\hspace*{-1cm}\textbf{Local complexity:}}}\\
\multicolumn{4}{l}{}\\
\textbf{model} & \textbf{time} 																				 & \textbf{probability} 			& \textbf{reference}	\\
\hline
\beep  & $O((\log \Delta + \log(1/\eps))\log (1/\eps))$ & $1-\eps$ 				& Thm.~\ref{thm:local-restate2} \\
\hline
\local & $O(\log \Delta + \log(1/\eps))$ & $1-\eps$ 									& \cite{G16}\\
\multicolumn{4}{l}{}\\
\multicolumn{4}{l}{{\hspace*{-1cm}\textbf{Global complexity:}}}\\
\multicolumn{4}{l}{}\\
\textbf{model} & \textbf{time} 																				& \textbf{probability} 			& \textbf{reference}	\\
\hline
\beep & $O(\log n)$																		& w.h.p. 			& \cite{feedback}\\
 			 & $O(\log^2 n)$														& w.h.p.			& \cite{beep13}\\

\hline
			 & $O(\log \Delta) + 2^{O(\sqrt{\log \log n})}$ & w.h.p. 			& \cite{G16}\\
\local & $O(\log n)$																	& w.h.p.			& \cite{Luby86}\\
 			 & $2^{O(\sqrt{\log n})}$													& 1						& \cite{DBLP:conf/stoc/PanconesiS92}\\
\end{tabular}
\end{center}
\caption{Overview of Results and Previous Work: Comparison of our results in the \beep model with results in the \local model and with previous work in the \beep and \local models. We include global bounds for completeness. While these global bounds yield w.h.p. guarantees, we want to stress that the key use of local analysis is when the success probability $1-\eps$ is small, e.g. constant. As demonstrated in~\cite{halldorsson2015local-podc}, w.h.p. improvements on global broadcast can be achieved using local bounds on MIS computations that are locally successful with very low probability. Also our $\Omega(\Delta)$ lower bound in Theorem~\ref{thm:local-BGG} on translating the Schneider Wattenhofer algorithm for Bounded Growth Graphs applies to global and local complexity, as this algorithm is deterministic, in which case these measures are the same.}
    \label{tab:truthTables}   
\end{table}

\section{Models and Definitions}
\textbf{\local and \beep Models:} In both models, the network is abstracted as an undirected graph $G=(V, E)$ where $|V|=n$. All nodes wake up simultaneously. Communication occurs in synchronous rounds. In the \local model (e.g.,~\cite{G16,peleg2000distributed}), each node knows its graph neighbors.  Nodes 
communicate reliably, where in each round nodes can 
exchange an arbitrary amount of information with their immediate graph neighbors.
On the other hand, in the \beep model (e.g.,~\cite{beep13,cornejo2010deploying}), nodes do not know their 
neighbors.  Nodes communicate reliably and a node can choose to either beep or listen. If a node $v$ listens in slot\footnote{To disambiguate, we refer to the rounds of the \beep model as \emph{slots}.} $t$ it can only distinguish between silence (no neighbor beeps in slot $t$) or the presence of one or more beeps (at least one neighbor beeps in in slot $t$).

\textbf{Graph-related Definitions:} 
We denote the set of \textit{$h$-hop neighbors} of node $v$ in $G$ by $N_h(v)=\{u\in V \mid d(u,v)\leq h\}$, where $d(u,v)$ indicates the \textit{hop-distance} between two nodes in a graph. By $\Delta := \max_{v \in V} |N_1(v)|-1$ we denote the \textit{maximum degree} of $G$. A set of vertices $I \subseteq V$ is an \textit{independent set} of $G$ if no two nodes in $I$ are neighbors in $G$. An independent set $I \subseteq V$ is a \textit{maximal independent set (MIS)} of $G$ if, for all $v \in V\setminus I$, the set $I\cup\{v\}$ is not independent. An event occurs \textit{with high probability} (w.h.p.), if it occurs with probability at least $1-n^{-c}$ for some constant $c \ge 1$.

\section{Algorithm}

We first review the MIS Algorithm of~\cite{G16} for the \local model and then describe our modification for the \beep model. 
\subsection{Algorithm of~\cite{G16} in the \local Model}\label{sec:alg-local}
The MIS algorithm of~\cite{G16} runs for 
$$R:=\beta(\log \degr + \log (2/\eps))=O(\log \degr + \log (1/\eps)) $$
 rounds, where $\beta=1300$. In each round $t$, each node $v$ has a \emph{desire-level} $p_t(v)$ for joining the MIS, which initially is set to $p_0(v)=1/2$. 
\begin{definition}[Effective Degree, Ghaffari~\cite{G16}]\label{def:effect}
The sum of the desire-levels of neighbors of $v$ is called its \emph{effective-degree} $d_{t}(v)$, i.e., $d_t(v)=\sum_{u \in N(v)} p_{t}(u)$. 
\end{definition}
The desire-levels change over time: 
$$p_{t+1}(v)= 
\left\{
\begin{tabular}{ll}
    $p_{t}(v)/2$, & if $d_{t}(v)\geq 2$,\\
    $\min\{2p_{t}(v), 1/2\}$, & if $d_{t}(v)< 2$
\end{tabular}\right. .
$$
The desire-levels are used as follows: 
In each round, node $v$ gets \emph{marked} with probability $p_{t}(v)$. If $v$ is marked, and no neighbor of $v$ is marked, $v$ joins the MIS and gets removed along with its neighbors. Using the power of the \local model, in each round $t$, nodes exchange exact values of $p_{t}(u)$ with all their neighbors.

\subsection{Our Algorithm in the \beep Model}\label{sec:beepalgo}

In emulating the MIS algorithm of Section~\ref{sec:alg-local} in the \beep model, we do not require that a node $v$ learn the exact values of $p_{t}(u)$ for all neighbors $u$ in order to compute $d_{t}(v)$. Instead, we allow node $v$ to decide, based on how many beeps $v$ receives within a certain number of rounds, whether $d_{t}(v)$ is more likely to be larger than $1/10$ or smaller than $22$. To estimate which of these two scenarios applies, node $v$ beeps with probability $p_{t}(v)$ for a certain number of times and counts how often it received a beep when it is not sending. The number of received beeps serves as an indicator to estimate whether $d_{t}(v)$ might be smaller than $22$ or larger than $1/10$. To perform this estimation, we define time intervals in the \beep model. Eventually, an sequence of two intervals and one additional time slot is used to emulate each round of the \local algorithm~\cite{G16} in the \beep model.
\begin{definition}[Interval of slots]
We define an \textit{interval} to consist of 
$$I:=2000(\ln(1500)+\ln(2/\eps))=O(\log(1/\eps))$$
 slots in the \beep model. 
\end{definition}
\textbf{During the first interval}, the algorithm computes the ratio of the number of beeps received ($b_t(v)$) to the 
total number of slots in which $v$ listened during the interval ($c_t(v)$) as follows: in each interval $t$, every node $v$ maintains two counters $c_t(v)$ and $b_t(v)$. Counter $c_t(v)$ counts the number of slots that $v$ is listening to the channel during interval $t$. Counter $b_t(v)$ counts the number of beeps $v$ receives during  interval $t$. Both counters $c_t(v)$ and $b_t(v)$ are initialized to $0$ at the beginning of interval $t$. In each of the $I$ slots of interval $t$, every node $v$ decides randomly to beep with probability $p_{t}(v)\leq 1/2$. In each slot where $v$ decides not to send, node $v$ listens to the channel and increases $c_t(v)$ by one. If $v$ receives a signal in this particular slot, node $v$ increases its counter $b_t(v)$ by one. After all $I$ time steps of interval $t$, node $v$ compares $c_t(v)$ and $b_t(v)$. In case $c_t(v) \leq I/3$ we assume node $v$ did not listen often enough to make an informed decision and let $v$ randomly choose whether $b_t(v)/c_t(v) > \frac{1}{5}$ or not with probability $1/2$ for each choice -- this is particular important when $c_t(v)=0$, as this avoids a division by $0$. If $c_t(v) > I/3$, node $v$ decides to update its desire-level:
\begin{eqnarray}
p_{t+1}(v)&=& 
\left\{
\begin{tabular}{ll}
    \vspace*{0.05cm}$p_{t}(v)/2,$ & if  $b_t(v)/c_t(v) > \frac{1}{5}$\\
    $\min\{2p_{t}(v), 1/2\},$  & if $b_t(v)/c_t(v) \leq \frac{1}{5}$
\end{tabular}\right..\label{eq:cond}
\end{eqnarray}
Thus, we replace the condition $d_{t}(v)\geq 2$ in the algorithm of~\cite{G16} by the condition $b_t(v)/c_t(v) > \frac{1}{5}$.
The ratio is chosen to be $1/5$, as in the analysis it turns out that this is a good ratio in order to decide whether the effective degree $d_{t}(v)$ is larger than $1/10$ or smaller than $22$. 

Notice that 1) these two ranges overlap, as we trade the uncertainty in making this decision for a shorter runtime while guaranteeing strong probabilities on correct decisions, and 2), that the overlap range $[1/10,22]$ is chosen to capture a safety-distance around $2$ that yields simple calculations in the proof.

For the sake of readability we replace $\beta$ used in the definition of the number of rounds $R$ in the \local algorithm in Section~\ref{sec:alg-local} by $\gamma$, which is the analogous to constant $\beta$ used in~\cite{G16} and above when we analyze the \beep algorithm. For the sake of simpler analysis, we set $\gamma:= 80\beta =104000$. 

While in Ghaffari's algorithm, in each round $t$, nodes exchange exact values of $p_{t}(u)$ with all their neighbors, we show how nodes estimate the value of $p_{t}(u)$.

\textbf{During the second interval}, a node decides whether to join a set $M$. Note that in Theorem~\ref{thm:local-restate2} we state that $M$ is locally an MIS with probability at least $1-\eps$. At the beginning of this second interval, a node $v$ gets \emph{marked} with probability $p_{t}(v)$ and does not change whether it is marked during the interval. If a node is marked in an interval, it selects half of the time slots in the interval uniformly at random and beeps in these time slots and listens in the others. If $v$ is marked, and does not receive a beep in those time slots where $v$ decides to listen, node $v$ concludes that none of its neighbors is beeping and thus none of its neighbors is marked marked, and $v$ joins $M$. 

\textbf{During the final time slot} that completes the emulation of a round of the \local algorithm, $v$ beeps to indicate it joined $M$. In this time slot, each node that beeps or receives a beep gets removed, which corresponds to removing all nodes in $M$ along with their neighbors in Ghaffari's MIS algorithm.

\section{Local Complexity of our MIS Algorithm}
We demonstrate that for each node $v$, the accuracy of deciding whether $v$'s effective degree is high or low is good enough for the translated algorithm of~\cite{G16} to work correctly and fast in the \beep model, i.e., our algorithm does not require $v$ to learn exact desire-values of its neighbors.
In Section~\ref{sec:good} we define good nodes as those nodes that estimate the effective degree (see Definition~\ref{def:effect}) accurately enough for our purposes, and bound the probability for a node being a good node in Lemma~\ref{lem:help1.2}. 
In Section~\ref{sec:change} we show that most of the time most nodes adjust their desire-values correctly in correspondence with the effective degree even they do not know its exact value. These Lemmas provide the tools for our modified analysis of~\cite{G16} in Section~\ref{sec:mod}.

\subsection{For Most Nodes, Effective Degrees are Classified Correctly}\label{sec:good}
We introduce the notion of good nodes in Definition~\ref{def:good}, which are essentially nodes that correctly classify whether their effective degree is high or low. We show that if node $v$ is good in interval $r$, node $v$ (i) draws correct conclusions about whether its effective degree is high or low, and (ii) adjusts its desire-values in the same way as in the algorithm of~\cite{G16}. The first statement follows directly from Definition~\ref{def:good} and the second materializes in the proof of Theorem~\ref{thm:local-restate2}. In Section~\ref{sec:mod}, these insights will allow us to modify the analysis of~\cite{G16} to obtain statements about good nodes. In order to apply this for a large part of the graph, we show that most nodes are good, which we use in Section~\ref{sec:mod} to obtain the desired statements about local complexity and correctness.

\begin{definition}[Good node]\label{def:good} A node $v$ is a \textit{good node} in an interval $t$, if at the end of the interval the following three conditions are satisfied: 
\begin{enumerate}
\item $c_t(v)> I/3$, and
\item If $b_t(v)/c_t(v) >    \frac{1}{5}$, then $d_{t}(v)\geq 1/10$, and\vspace*{0.2mm}
\item If $b_t(u)/c_t(v) \leq \frac{1}{5}$, then $d_{t}(v)\leq 22$.
\end{enumerate}
\end{definition}
The main result of this section is that many nodes are good, which is formalized as follows:
\begin{lemma}\label{lem:help1.2}
For any node $v$ and interval $t$, the probability that $v$ is a good node is at least $1-2e^{-I/100}$. 
\end{lemma}
We prove Lemma~\ref{lem:help1.2} at the end of this Subsection. To prepare for the proof, we introduce two sub-Lemmas, Lemma~\ref{lem:help1.1} and~\ref{lem:help2}, to bound the probability that $b_t(v)/c_t(v)$ reflects whether the effective degree is high or low based on the condition $b_t(v)/c_t(v) > \frac{1}{5}$ that we use rather than $d_t(v)\geq 2$ that is used by~\cite{G16}. This differs from~\cite{G16}, as in the \local model, full information on neighbor's effective degrees can be obtained within one round of communication, while we can only operate with beeps. These Lemmas use the Chernoff Bound stated in 
\ifshort
Lemma~\ref{FULL:lem:chernoff}
\fi
\iffull
Lemma~\ref{lem:chernoff}
\fi
 in the Appendix for completeness. 

The following Lemma states that for any node $v$, most of the time property 1. in Definition~\ref{def:good} is satisfied, such that $v$ listens often enough to make an informed decision.
\begin{lemma}\label{lem:help1.1}
For any node $v$ and interval $t$, $Pr(c_t(v) > I/3) > 1-e^{-I/36}$. 
\end{lemma}
\ifshort
The proof of the Lemma uses a standard application of Chernoff Bounds, see 
\ifshort
Lemma~\ref{FULL:lem:chernoff}
\fi
\iffull
Lemma~\ref{lem:chernoff}
\fi
in the Appendix. We state the proof in the full version of the paper, see Appendix, Lemma~\ref{FULL:lem:help1.1}. 
\fi
\iffull
\begin{proof}
In each of the first $I$ slots of interval $t$, the probability that $v$ is not listening is $p_t(v)$, which is upper bounded by $1/2$ in the definition of our algorithm and we conclude $\mathbb{E}[c_t(v)]\geq I/2$. We apply Chernoff Bound 2. of 
\ifshort
Lemma~\ref{FULL:lem:chernoff}
\fi
\iffull
Lemma~\ref{lem:chernoff}
\fi
 (see Appendix) for $X=c_t(v)$ with $\delta:=1/3$ and obtain that $Pr(c_t(v)\leq I/3)\leq e^{-I/36}$. 
\end{proof}
\fi

The next Lemma states that for any node $v$, most of the time property 2. or 3. in Definition~\ref{def:good} are satisfied. 
\begin{lemma}\label{lem:help2}
Assume $c_t(v) > I/3$ and let $b_t(v)/c_t(v)$ be the ratio computed by node $v$ in an interval.
\begin{enumerate}
\item If $b_t(v)/c_t(v) > \frac{1}{5}$, then $Pr(d_{t}(v)\geq 1/10) \geq 1-e^{-I/100}$, and
\item if $b_t(v)/c_t(v) \leq    \frac{1}{5}$, then $Pr(d_{t}(v)\leq 22) \geq 1-e^{-I/100}$.
\end{enumerate}
\end{lemma} 
Notice that although $d_t(v)$ changes its value over time, we can bound $d_t(v)$ at time $t$ with the probabilities stated in this Lemma independent of the history of $d_t(v)$. 
\ifshort
In this submission we only sketch the proof due to space reasons and state the full proof in the full version of the paper, see Appendix, Lemma~\ref{FULL:lem:help2}. Sketch of proof of Lemma~\ref{lem:help2}: we define a random variable for each time slot that indicates whether any neighbor of $v$ beeps. The probability $q_t(v)$ that any neighbor of $v$ beeps is the same for each time slot within the same interval $t$. Using Chernoff Bounds, we bound the probability that $q_t(v)\geq 1/10$ or $q_t(v)\leq 2/5$ depending on whether $b_t(v)/c_t(v) > \frac{1}{5}$ or $b_t(v)/c_t(v) \leq \frac{1}{5}$. The value $q_t(v)$ then can be used to upper and lower bound $d_t(v)$, which is the probability mass distribution among the neighbors of $v$ that ultimately determined $q_t(v)$ in the first place. By doing so, we derive the stated bounds on $d_t(v)$. 
\fi
\iffull
\begin{proof}

For each time slot $i$ of interval $t$, let $q_t(v)$ be the probability that at least one of the
neighbors of node $v$ beeps during slot $i$ of interval $t$. Note that for each time slot $i$, the probability $q_t(v)$ is the same and choices are made for each $i$ independently. In the following, for each $t$, we consider $c_t(v)$ to be fixed evaluations of the random variable describing them. Define independent random variables $X_1,\dots,X_{c_t(v)}$ for each of the $c_t(v)$ time slots in interval $t$ during which $v$ was listening, where $X_i=0$ or $1$ indicates whether $v$ received a \beep during the $i$'th slot in which $v$ listened. Define $X:=\sum_{i=1}^{c_t(v)}X_i$, the random variable that indicates how many beeps $v$ receives during interval $t$. We conclude that $\mathbb{E}[X]=q_t(v)\cdot c_t(v)$. Now observe that $b_t(v)$ is the sum of $c_t(v)$ evaluated random variables $X_i$, and thus an evaluation of $X$. 

\textit{Proof of Statement 1:} Assume $b_t(v)/c_t(v) > \frac{1}{5}$. We first show that in this case $Pr[q_t(v)\geq 1/5]\geq 1-e^{-I/66}$ and then derive the claimed statement on $d_t(v)$. Observe that the Bernoulli distribution $X$ is defined using $q_t(v)$ and $q_t(v)$ only, and $Pr[X/c_t(v)\leq 1/5]$ is a monotonically increasing function of $q_t(v)$ and vice versa. This allows us to turn the analysis around and analyze the probability that an evaluation of $X$ is larger than $c_t(v)/5$ given $q_t(v)$, and draw conclusions on $q_t(v)$ from the event $X$ is larger than $c_t(v)/5$. Therefore, we now assume for this part of the proof that $q_t(v)\leq 1/10$. In this case we know that $Pr[X/c_t(v)\geq 1/5] = Pr[X/c_t(v)\geq (1+(\frac{1}{5q_t(v)}-1))q_t(v)c_t(v)]$.
. We apply Chernoff Bound 1.~of 
\ifshort
Lemma~\ref{FULL:lem:chernoff}
\fi
\iffull
Lemma~\ref{lem:chernoff}
\fi
 (see Appendix) with $\delta=\frac{1}{5q_t(v)}-1$, which we can do due to the assumption of $q_t(v)\leq 1/10$, and obtain that 
\begin{eqnarray*}
Pr[X/c_t(v)\geq (1+(\frac{1}{5q_t(v)}-1))q_t(v)c_t(v)]
&=&
Pr[X\leq (1+\delta)\mathbb{E}[X]]
\\
&\leq&
e^{-\frac{\delta^2}{2+\delta}\mathbb{E}[X]}
\\
&=&
e^{-\frac{(\frac{1}{5q_t(v)}-1)^2}{2+(\frac{1}{5q_t(v)}-1)}q_t(v)c_t(v)}
\\
&=&
e^{-\frac{(\frac{1}{5}-2q_t(v)+5q_t(v)^2)}{1+5q_t(v)}c_t(v)}
\\
&\leq&
e^{-\frac{c}{22}}
\\
&\leq&
e^{-I/66}
\end{eqnarray*}
due to assuming $q_t(v)\leq 1/10$ and $c_t(v) > I/3$ (see statement of the Lemma). From this we derive that if $X\geq c_t(v)/5$, then $Pr[q_t(v)\geq 1/10]\geq 1-e^{-I/60}$.
Now we lower bound $d_t(v)$ based on $q_t(v)$. We know that $q_t(v)=1-\prod_{u \in N(v)} (1-p_t(u))$ and $d_t(u)=\sum{u \in N(v)} p_t(u)$. As the $d_t(v)$ is minimized when the whole probability mass of $q_t(v)$ is aggregated in one node, we conclude that $Pr[d_t(v)\geq 1/10]\geq 1-e^{-I/66}\geq 1-e^{-I/60}$.

\textit{Proof of Statement 2:} Assume $b_t(v)/c_t(v) \leq \frac{1}{5}$. In the following proof, for each time slot of interval $t$, we lower bound the probability that the distribution-probability $q_t(v)$ is smaller than $1/2$. Then we upper bound $d_t(v)$ based on $q_t(v)$. To be more specific, we first show that in case an evaluation of $X$ is smaller than $c_t(v)/5$ it is $Pr[q_t(v)\leq 1/2]\geq 1-e^{-I/100}$. 

As argued above, we can turn the analysis around and assume $q_t(v)> 1/2$ and show in the text below, that under this assumption, $Pr[X/c_t(v) > \frac{1}{5}]\geq 1-e^{-I/90}$ and in turn draw conclusions on the probability of $q_t(v)>1/2$ when $X/c_t(v) > \frac{1}{5}$ is given. Assuming $q_t(v)> 1/2$, we derive $Pr[X/c_t(v) > \frac{1}{5}]=Pr[X\leq c_t(v)/5]\leq Pr[X\leq 2/5q_t(v)c_t(v)]=$ when $Pr[X\leq 2/5\mathbb{E}[X]]$, as $\mathbb{E}[X]=q_t(v)c_t(v)$.
We apply Chernoff Bound 1.~of 
\ifshort
Lemma~\ref{FULL:lem:chernoff}
\fi
\iffull
Lemma~\ref{lem:chernoff}
\fi
 (see Appendix) to upper bound the probability that an evaluation of the random variable $X$ is larger than $2$ times its expectation, which we achieve with $\delta=3/5$
\begin{eqnarray*}
Pr[X\leq 2/5\mathbb{E}[X]]
&=&
Pr[X\leq (1-\delta)\mathbb{E}[X]]
\\
&\leq&
e^{-\frac{\delta^2}{2+\delta}\mathbb{E}[X]}
\\
&=&
e^{-\frac{9}{65}\cdot q_t(v)c_t(v)}
\end{eqnarray*}
Using the assumption that $q_t(v)>1/2$ and the assumption that $c_t(v) > I/3$ (see Lemma statement), we can bound this further by $e^{-3I/130}\leq e^{-I/100}$. Now we know that if $q_t(v)>1/2$, then $Pr[X/c_t(v) \leq 1/5]\leq e^{-I/100}$. We conclude that if $X/c_t(v) \leq 1/5$, then $Pr[q_t(v)>1/2]\leq e^{-I/100}$, as $X$ is defined using $q_t(v)$ and $q_t(v)$ only, and defined in a way that the implied Bernoulli distribution monotonically increases when $q_t(x)$ increases. From this we derive that if $X/c_t(v) \leq 1/5$, then $Pr[q_t(v)\leq 1/2]\geq 1- e^{-I/100}$

Next, we know that $q_t(v)=1-\prod_{u \in N(v)} (1-p_t(u))$ and $d_t(u)=\sum_{u \in N(v)} p_t(u)$. As the whole probability mass of $q_t(v)$ could be distributed evenly among $v$'s neighbors, the worst case is $q_t(v)=1-(1-p_t)^\degr\geq 1-e^{-\degr/p_t}$ with $p_t(u)=p_t$ for any $u\in N(v)$. From this we conclude that $1/2\geq 1-e^{-\degr/p_t}$, which in turn yields $p_t\leq - \degr/\log(1/2)\leq 5\degr < 22\degr$.

Due to $q_t(v)\leq 1/2$ with probability $1-e^{-I/100}$, we conclude $p_t/\degr\leq 22$. The definition of $d_t(v)$ yields $d_t(v)\leq 22$. From this we conclude $Pr[d_t(v)\leq 22]\geq 1-e^{-I}\geq 1-e^{-I/100}$.
\end{proof}
\fi

Now we are ready to prove Lemma~\ref{lem:help1.2}, which follows from combining Lemma~\ref{lem:help1.1} and Lemma~\ref{lem:help2} to cover all properties of Definition~\ref{def:good} and multiplying the probabilities of the related events stated in the Lemmas we use. We state the full proof in the full version of the paper, see Appendix, Lemma~\ref{lem:help1.2}. 
\iffull
\begin{proof}(of Lemma~\ref{lem:help1.2}).
Due to Lemma~\ref{lem:help1.1}, we know that $Pr(c_t(u) > I/3) > 1- e^{-I/36}$, such that property (1) of a good node (Definition~\ref{def:good}) is satisfied. Now we can assume $c_t(u)> I/3$ with probability $1- e^{-I/36}$, then the probability that properties (2) and (3) of a good node are satisfied is at least $1-e^{-I/100}$ each due to Lemma~\ref{lem:help2}. We conclude that all three conditions are satisfied for node $u$ in an interval with probability larger than
$\left(1-e^{-I/36}\right)\left(1-e^{-I/100}\right)>1-2e^{-I/100}.$
Finally, notice that $b_t(v)$ is a realization of the random variable $X$.
\end{proof} 
\fi

\subsection{Changes of Effective Degrees Based on Neighbor's Behavior}\label{sec:change}
We show in Lemma~\ref{lem:help1.3} that in any interval, the effective degree of a node $v$ (see Definition~\ref{def:effect}), that is contributed by a set of neighbors with high effective degree in that interval, shrinks by almost a factor of $2$ with significant probability. This is a key part in the modification of the analysis of~\cite{G16} in Section~\ref{sec:mod}. This is stated in a formal way in Lemma~\ref{lem:help1.3} and proven using Lemma~\ref{lem:help1.2}.   

The following Lemma is a key Lemma. It allows us to bound the amount of $d_{t+1}(v)$ that is contributed by neighbors of $v$ with high effective degree based on the amount of $d_{t}(v)$ that is contributed by neighbors of $v$ with high effective degree. A similar bound is used in~\cite{G16}, where it is obtained in a straight forward way thanks to the power of the \local model. We need to (and already did) work a bit harder to obtain a similarly useful bound. This Lemma shows that the precise way of increasing/reducing the desire value of a node based on $d_t(v)$, which requires full knowledge of all these values of neighbors of $v$ in~\cite{G16}, can be replaced by estimating $d_{t}(v)$ using beeps. Of course the bound is less strong as in~\cite{G16}, we obtain only a decrease of $51/100$ vs. $1/2$ in~\cite{G16}. Also, as we operate probabilistically, we can only claim this bound with a certain probability. Fortunately this probability is sufficiently high and the ratio $51/100$ strong enough to allow us to modify the analysis of~\cite{G16} correspondingly, as we do in the rest of this section.  

\begin{lemma}\label{lem:help1.3}
For any interval $t$ and node $v$, it is the case that $d_{t+1}(v|N^{\geq 22}_{1,t}(v))\leq \frac{51}{100}d_t(v|N^{\geq 22}_{1,t}(v))$ with probability at least $1-300e^{-I/2000}$.
\end{lemma}
The proof of this Lemma splits up the amount of $d_t(v)$ that is contributed by nodes with high effective degree into two parts. One part is contributed by good nodes, the other one by bad nodes. We mainly need to work to keep the contribution of bad nodes in check, as they may increase their desire values when they shouldn't (but they don't know). This can be done using Lemma~\ref{lem:help1.2} and a Chernoff Bound. However, problems arise when the set of high effective degree neighbors is small, smaller than $100I$ to be precise. In this case the probabilities that we obtain we Chernoff are not strong enough to modify the analysis of~\cite{G16}, i.e., not negatively exponential in $I$, and we treat this case of less than $100I$ nodes separately.
The following notation helps us to formalizes some of the above:
\begin{definition}[$N^{\geq 22}_{1,t}(v)$, $\bar{N}^{\geq 22}_{1,t}(v)$ and $d_{t}(v|S)$]
Denote by $N^{\geq 22}_{1,t}(v)=\{u\in N_1(v) | d_t(u)\geq 22\}$ the neighbors of $v$ with $d_t(u)\geq 22$. Denote by $\bar{N}^{\geq 22}_{1,t}(v)=\{u\in N^{\geq 22}_{1,t}(v) | u$ is not good$\}$ the set of nodes in $N^{\geq 22}_{1,t}(v)$ that are not good. Let $S\subseteq N_1(v)$ be a set of nodes, then we denote by $d_{t}(v|S)=\sum_{u\in S}p_t(u)$ the amount of $d_t(v)$ contributed by nodes in $S$. 
\end{definition}

\begin{proof}(of Lemma~\ref{lem:help1.3}). 
Based on how node $v$ adjusts its value $p_t(v)$ when executing the algorithm (see Equation~\ref{eq:cond} that depends on the $b_t(v)/c_t(v)$ ratio), we can bound $d_{t+1}(v|N^{\geq 22}_{1,t}(v))$ to be smaller than 
$\frac{1}{2}d_t(v|N^{\geq 22}_{1,t}(v)\setminus \bar{N}^{\geq 22}_{1,t}(v)) + 2d_t(v|\bar{N}^{\geq 22}_{1,t}(v))$
 and this is derived directly from the definition of the algorithm and in correspondence to how we modify the Algorithm of Ghaffari. Now we can write $\frac{1}{2}d_t(v|N^{\geq 22}_{1,t}(v)\setminus \bar{N}^{\geq 22}_{1,t}(v))$ as $\frac{1}{2}d_t(v|N^{\geq 22}_{1,t}(v)) - \frac{1}{2}d_t(\bar{N}^{\geq 22}_{1,t}(v))$ and when applied to the previous bound derive that 
$d_{t+1}(v|N^{\geq 22}_{1,t}(v)) \leq \frac{1}{2}d_t(v|N^{\geq 22}_{1,t}(v)) + \frac{3}{2}d_t(v|\bar{N}^{\geq 22}_{1,t}(v)).$ 
To analyze the probability, that this is at most $\frac{51}{100} d_t(v|N^{\geq 22}_{1,t}(v))$, we distinguish two cases. In case 1, we consider $|N^{\geq 22}_{1,t}(v)|< 100I$ and in case 2 we consider $|N^{\geq 22}_{1,t}(v)| \geq 100I$.
\\
\textbf{Case 1, $|N^{\geq 22}_{1,t}(v)|< 100I$:}
The probability that no node in $N^{\geq 22}_{1,t}(v)$ is bad, i.e., $\bar{N}^{\geq 22}_{1,t}(v)=\emptyset$ is $(1-2e^{-I/100})^{|N^{\geq 22}_{1,t}(v)|}$, due to Lemma~\ref{lem:help1.2}. By the assumption of case 1, that there are at most $I$ neighbors $u$ of $v$ with $d_t(u)\geq 22$, this can be bounded by $\geq (1-2e^{-I/100})^{100I}$, which in turn is larger than $1-100I\cdot 2e^{-I/100}=1-200e^{-I/100+\log(I)}\geq 1-300e^{-I/100}\geq 1-300e^{-I/2000}$ due to the choice of $I$.
\\
\textbf{Case 2, $|N^{\geq 22}_{1,t}(v)|\geq 100I$:}
We bound the probability that at most a $1/150$ fraction of the nodes in $N^{\geq 22}_{1,t}(v)$ is bad, i.e., $|\bar{N}^{\geq 22}_{1,t}(v)|\leq |N^{\geq 22}_{1,t}(v)|/150$. To do so, we apply Chernoff Bound 2 of 
\ifshort
Lemma~\ref{FULL:lem:chernoff}
\fi
\iffull
Lemma~\ref{lem:chernoff}
\fi
, see Appendix, for $X_i=$ node $v_i \in \{v_1,\dots,v_{|N^{\geq 22}_{1,t}(v)|}\}=N^{\geq 22}_{1,t}(v)$ is good. Based on Lemma~\ref{lem:help1.2}, we can conclude that $\mathbb{E}[X]\geq (1-2e^{-I/100})\cdot|N^{\geq 22}_{1,t}(v)|$. Choosing $\delta=1/300$, we can bound 
\begin{eqnarray*}
&&
Pr[X\leq (1-1/150)\cdot|N^{\geq 22}_{1,t}(v)|]
\leq 
Pr[X\leq (1-1/300)(1-2e^{-I/100})\cdot|N^{\geq 22}_{1,t}(v)|]
\\
&&
=
Pr[X\leq (1-\delta)\mathbb{E}[X]]
=
e^{-\frac{\delta^2}{2}\mathbb{E}[X]}
=
e^{-\frac{1}{180000}\mathbb{E}[X]}
\leq
e^{-\frac{1}{180000}(1-2e^{-I/100})\cdot|N^{\geq 22}_{1,t}(v)|}
\end{eqnarray*}
Now we use the assumption $d_t(v|N^{\geq 22}_{1,t}(v))$ and the definition of $I$ to derive that this is smaller than
$
e^{-\frac{1}{180000}(1-2e^{-I/100})\cdot 100I}
\leq
e^{-I/2000}
\leq
300 e^{-I/2000}.
$
From this we conclude that $Pr[|\bar{N}^{\geq 22}_{1,t}(v)|\leq |N^{\geq 22}_{1,t}(v)|/150]\geq 1-300 e^{-I/2000}$. 

When combining both cases 1 and 2, we obtain that with probability at least $1- 300e^{-I/2000}$, the value of $d_{t+1}(v|N^{\geq 22}_{1,t}(v))$ is smaller than 
$\frac{50}{100}d_t(v|N^{\geq 22}_{1,t}(v)) + \frac{3}{2}\frac{1}{150}d_t(v|N^{\geq 22}_{1,t}(v)) = \frac{51}{100}d_t(v|N^{\geq 22}_{1,t}(v)).$
\end{proof} 

\subsection{Proof of Theorem~\ref{thm:local-restate2}}\label{sec:mod}

Now we are prepared to follow the analysis of~\cite{G16} and adapt it to our modifications of the algorithm. Using the notation used in the last two sections, Theorem~\ref{thm:local-correctness} and~\ref{thm:local-restate2} are derived from:
\begin{theorem} \label{thm:local-restate} For each node $v$, the probability that $v$ makes a (locally correct) decision within the first $R$ intervals is at least $1-\eps$. 
Furthermore, this holds even if the outcome of the coin tosses outside $N^{+}_{2}(v)$ are determined adversarially.
\end{theorem}
The rest of this Section is devoted to proving Theorem~\ref{thm:local-restate}. First we define two kinds of \emph{golden intervals} for a node $v$, by analogy with the definition of \emph{golden rounds} in~\cite{G16}, then we show that it is likely that there are many golden intervals in case a node does not join $M$ (Lemma~\ref{lem:goldCount}). Then we argue that, if there are that many golden intervals, then it is likely that a node gets removed due to either joining $M$ or having a neighbor that joins $M$ (Lemma~\ref{lem:6}). To prove Lemma~\ref{lem:goldCount} we use Lemma~\ref{lem:help1.3}; to prove Lemma~\ref{lem:6} we use Lemma~\ref{lem:help1.2}.
\begin{definition}
A node $v$ has \emph{likely-low effective degree} if $b_t(v)/c_t(v) \leq \frac{1}{5}$, and has \emph{likely-high effective degree} if $b_t(v)/c_t(v) > \frac{1}{5}$. 
\end{definition}
\begin{definition}[Golden intervals]
Interval $t$ is a golden interval of type 1, if $b_t(v)/c_t(v) \leq \frac{1}{5}$ and $p_{t}(v)= 1/2$. Interval $t$ is a golden interval of type 2 if $b_t(v)/c_t(v) > \frac{1}{5}$ and at least $d_{t}(v)/11$ of $d_{t}(v)$ is contributed by neighbors $u$ with $d_{t}(u)\leq 22$ (nodes of \textit{low effective degree}). 
\end{definition}

These are called golden intervals because, as we will see, in the first type, $v$ has a constant chance of joining $M$ and in the second type, there is a constant chance that one of those neighbors of $v$ with low effective degree joins $M$ and thus $v$ gets removed. 

The following lemma and proof follow along the lines of a similar proof in~\cite{G16}, for Theorem 3.1, and is modified to our setting using the Lemmas proven so far.

\begin{lemma}\label{lem:goldCount} By the end of interval $R$, with probability at least $1-1500e^{-I/2000}$, either $v$ has joined, or has a neighbor in $M$, or at least one of its golden interval counts reached $R/13$.
\end{lemma}
\ifshort
In this submission we only sketch the proof due to space reasons and state the proof in the full version of the paper, see Appendix, Lemma~\ref{FULL:lem:goldCount}. Sketch of proof: We follow the lines of the proof of Lemma 3.2 in~\cite{G16}, while modifying parameters to match ours and take care of the probabilistic behavior our algorithm introduces. The main parts of treating probabilities are achieved by replacing a bound~\cite{G16} applies, which is immediate in the \local model, with our Lemma~\ref{lem:help1.3} and applying basic tools such as Markov Bounds.
\fi
\iffull
\begin{proof} Let $g_1$ and $g_2$ respectively be the number of golden intervals of types 1 and 2 for $v$ during this period.  We assume, that by the end of interval $R$, node $v$ is not removed and $g_1 \leq R/13$. Otherwise the statement of the Lemma would already be satisfied. Based on this assumption, we lower bound in the remaining part of this proof the number $g_2$ of golden intervals of type-2 while taking into account that any node $u$'s ratio $b_t(u)/c_t(u)$ might not always correctly represent whether $d_t(u)\geq 2/5$ or $d_t(u)\leq 22$.

Let $h$ be the number of intervals where $b_t(v)/c_t(v) > \frac{1}{5}$. Notice that the changes in $p_{t}(v)$ are governed by the condition $b_t(v)/c_t(v) > \frac{1}{5}$ and intervals with $b_t(v)/c_t(v) > \frac{1}{5}$ are exactly the ones in which $p_{t}(v)$ decreases by a $2$ factor. Since the number of $2$ factor increases in $p_{t}(v)$ can be at most equal to the number of $2$ factor decreases in it, we get that there are at least $R-2h$    intervals in which $p_{t}(v)=1/2$. 

Now out of these $g_1>R-2h$ intervals, at most $h$ of them can be when $b_t(v)/c_t(v) >\frac{1}{5}$. Hence, $g_1 \geq R-3h$. As we have assumed $g_1 \leq R/13$, we get that $R-3h \leq R/13$, and conclude that $h\geq R\cdot 4/13$.

Let us consider the changes in the effective-degree $d_{t}(v)$ of $v$ over time. Note that $d_{t}(v)$ reflects all changes of each neighbor $u$'s value $p_t(u)$ based on whether $b_{t'}(v)/c_{t'}(v) > \frac{1}{5}$ in previous intervals $t'<t$. This is independent of the actual value of $d_{t'}(v)$ at that time and thus $d_{t}(v)$ takes all previously made errors into account. 

If $b_t(v)/c_t(v) > \frac{1}{5}$ and this is not a golden interval of type-2, then we know that at most $\frac{1}{11} d_t(v)$ of $d_t(v)$ is contributed by neighbors $u$ with low effective degree $d_t(u)\leq 22$, such that the fraction of $d_t(v)$ contributed by those nodes doubles at most. On the other hand, at most all $d_t(v)$ of $d_t(v)$ is contributed by neighbors $u$ with high effective degree $d_t(u)\geq 22$, i.e., $d_{t}(v|N^{\geq 22}_{1,t}(v))\leq d_t(v)$. Due to Lemma~\ref{lem:help1.3} we know that $d_{t+1}(v|N^{\geq 22}_{1,t}(v))\leq \frac{51}{100}d_t(v|N^{\geq 22}_{1,t}(v))\leq\frac{51}{100} d_{t}(v)$ with probability at least $1-300e^{-I/2000}$. From this we conclude that with probability at least $1-300e^{-I/2000}$, 
$d_{t+1}(v) \leq 2 \frac{1}{11} d_t(v)+ d_{t+1}(v|N^{\geq 22}_{1,t}(v)) \leq \frac{765}{1100} d_{t}(v) < \frac{7}{10} d_{t}(v)$
There are $g_2$ golden intervals of type-2. We just showed that for all intervals with $b_t(v)/c_t(v) > \frac{1}{5}$, that are not among these $g_2$ golden intervals, the effective-degree $d_{t}(v)$ shrinks by at least a $7/10$ factor with probability at least $1-300e^{-I/2000}$ and this is independent of whether $b_t(v)/c_t(v) > \frac{1}{5}$ indicates the correct range of $d_t(v)$. Now let $g$ be the number of intervals with $b_t(v)/c_t(v) > \frac{1}{5}$, that are not among these $g_2$ golden intervals. We show that with probability at least $1-1500e^{-I/2000}$ in at least $4g/5$ of these intervals the effective-degree $d_{t}(v)$ shrinks by at least a $7/10$ factor. 
Let $k$ be the number of intervals in which the effective-degree $d_{t}(v)$ does not shrink by at least a $7/10$ factor, then $\mathbb{E}[k]\leq 300e^{-I/2000}g$. Using Markov's Inequality yields that
$Pr(k\geq g/5)\leq \frac{\mathbb{E}[k]}{g/5}\leq \frac{300e^{-I/2000}g}{g/5}\leq 1500e^{-I/2000}$

In the $g_2$ golden intervals of type-2 and the $g/5$ intervals in which $b_t(v)/c_t(v) > \frac{1}{5}$ that are not golden intervals of type-2 and in which the effective-degree $d_{t}(v)$ does not shrink by at least a $7/10$ factor, the value of $d_t(v)$ increases by at most a $2$ factor. Each of these $g_2+g/5$ intervals cancels the effect of at most $2$ shrinkage intervals, as $(7/10)^2 \times 2 < 1$. Thus, ignoring the total of at most $3(g_2+g/26)\leq 3(g_2+R/26)$ intervals lost due to type-2 golden intervals and their cancellation effects, every other interval with $b_t(v)/c_t(v) > \frac{1}{5}$ pushes the effective-degree of $v$ down by a $2/3$ factor.
 This cannot (continue to) happen more than $\log_{3/2} \degr$ times, as that would lead the effective degree to exit the $d_{t}(v)\geq 1/10$ region for any node. Hence, the number of intervals in which $b_t(v)/c_t(v) > \frac{1}{5}$ is at most $\log_{3/2} \degr + 3(g_2+g/5)$ with probability at least $1-1500e^{-I/2000}$. That is, $h \leq \log_{3/2} \degr + 3(g_2+g/5)$ with probability at least $1-1500e^{-I/2000}$. Since $h\geq R4/13$, we get $g_2 >R/26$ with probability at least $1-1500e^{-I/2000}$ due to the definition of $R$.

\end{proof}
\fi

The following Lemma is adapted from a proof of Lemma 3.3 of~\cite{G16} based on the new values and thresholds used in the modified algorithm and the corresponding definitions we introduced in this paper, and takes the error source and probabilistic behavior into account, which we generate due to not communicating effective degrees explicitly when (compared to what Ghaffari~\cite{G16} does in the \local model).

\begin{lemma}\label{lem:6}  In each type-1 (resp., type-2) golden interval, with probability at least $1/2000$, $v$ joins $M$ (resp., one of $v$'s neighbors joins $M$). If $R/13$ intervals are golden, then the probability that $v$ has not decided whether it is in $M$ during the first $R$ intervals is at most $\eps/2$. These statements hold even if the coin tosses in $N^{+}_{2}(v)$ are determined adversarially.
\end{lemma}
\ifshort
In this submission we only sketch the proof due to space reasons and state the proof in the full version of the paper, see Appendix, Lemma~\ref{FULL:lem:6}. Sketch of proof: We mainly use Lemma~\ref{lem:help2} and Lemma~\ref{lem:help1.3}, and enjoy the way parameters are chosen throughout this paper to proof this Lemma.
\fi
\iffull
\begin{proof}
In each type-1 golden interval, node $v$ gets marked with probability $1/2$. In such an interval it is the case that $b_t(v)/c_t(v) \leq \frac{1}{5}$ and therefore $d_{t}(v)\leq 3$ with probability at least $1-2e^{-I/100}$ by Lemma~\ref{lem:help1.2}. We conclude that the probability that no neighbor of $v$ is marked is 
\begin{eqnarray*}
\left(1-2e^{-I/100}\right)\cdot\prod_{u \in N(v)} \left(1-p_t(u)\right)
&\geq& \left(1-2e^{-I/100}\right)\cdot 4^{-\sum_{u \in N(v)} p_t(v)} \\
&=& \left(1-2e^{-I/100}\right)\cdot 4^{-d_{t}(v)} \\
&\geq& \left(1-2e^{-I/100}\right)\cdot 4^{-3}>1/100.
\end{eqnarray*}
Hence, $v$ joins $M$ with probability at least $1/100\cdot 1/2 >1/2000$.

Now consider a type-2 golden interval. In such an interval it is $b_t(v)/c_t(v) > \frac{1}{5}$ and due to Lemma~\ref{lem:help1.2} we know that $d_{t}(v)\geq 1/10$ with probability at least $(1-2e^{-I/100})$. For the sake of analyis, suppose we walk over the set $L$ of low effective degree neighbors of $v$ one by one and expose their randomness until we reach a node that is marked. We will find a marked node with probability at least 
\begin{eqnarray*}
\left(1-2e^{-I/100}\right)\cdot\left(1-\prod_{u \in \textit{L}} (1-p_{u}(t))\right) 
&\geq&
\left(1-2e^{-I/100}\right)\cdot\left(1- e^{-\sum_{u \in \textit{L}} p_{u}(t)}\right)\\
&\geq& \left(1-2e^{-I/100}\right)\cdot\left(1-e^{-d_{t}(v)/11}\right)\\
&\geq& \left(1-2e^{-I/100}\right)\cdot\left(1-e^{-1/110}\right)\\
& > &\left(1-2e^{-I/100}\right)\cdot 0.009 > 0.008,
\end{eqnarray*} 
where the last bound is due to choice of $I\geq 2000\cdot\log 1500$.
When we reach the first low effective degree neighbor $u$ that 1) satisfies the condition $d_{t}(u)\leq 22$ of $v$'s type-2 golden interval, and 2) that is marked, then the probability that no neighbor of $u$ gets marked is at least 
$$\prod_{w\in N(u)} (1-p_{t}(w)) \geq 4^{-\sum_{w \in N(u)} p_t(w)} \geq 4^{-d_{t}(u)} \geq 1/64.$$ Hence, with probability at least $0.008/64=1/8000$, one of the neighbors of $v$ joins $M$. 

We now know that in each golden interval, $v$ gets removed with probability at least $1/8000$, due to joining $M$ or having a neighbor join $M$. Thus, using Lemma~\ref{lem:goldCount}, we get that the probability that $v$ does not get removed is at most 
$$(1-1/8000)^{R/13}=(1-1/8000)^{8000(\log \degr + \log (2/\eps))} \leq \eps/(2\degr) \leq \eps/2$$
 due to the choice of $\gamma=104000$ in the definition of $R$ at the end of Section~\ref{sec:beepalgo}.
\end{proof}
\fi
Finally we are ready to prove Theorem~\ref{thm:local-restate}.
\begin{proof}(of Theorem~\ref{thm:local-restate}). 
Due to Lemma~\ref{lem:goldCount}, by the end of interval $R$, with probability at least $1-1500e^{-I/2000}$, either $v$ has joined, or has a neighbor in $M$, or at least one of its golden interval counts reached $R/13$. In the latter case, we know due to Lemma~\ref{lem:6}, that with probability at least $1-\eps/2$, a node $v$ terminates within $R$ intervals and decides whether it is in $M$. Therefore, the probability that a node terminates within $R$ intervals and decides whether it is in $M$ is at least $1-\eps/2$, as we chose $I=2000(\ln(1500)+\ln(2/\eps))$.

Finally we need to analyze local correctness of the computation, which is handled in the second intervals of each emulated rounds of the \local algorithm. The analysis above based on Lemma~\ref{lem:goldCount} argues that a node joins $M$ with a certain probability or gets removed due to a neighbor joining $M$. This satisfies the maximality condition of an MIS - no node could be added to $M$ without violating independence. Now we argue that independence is guaranteed locally with probability at least $1-\eps\frac{2^{-1000}}{\degr}$, i.e., no two neighboring nodes join $M$. As neighbors of a node $v$ that joins the $M$ are getting removed, we only analyze the probability that two neighbors join $M$ in the same round (in which case the MIS condition is violated locally). This can only happen if: given a node $v$, a subset $U$ of $v$'s neighbors marked themselves and all of them chose the same subset of time slots in the second interval to beep as $v$ did, such that none of them recognizes that the other node marked itself as well. Furthermore, to ensure that there is a neighbor $u$ in $u$ that indeed joins $M$, node $u$ must not have neighbors besides $v$ that marked themselves, or all of $u$'s neighbors that marked themselves also chose the same subset of time slots to beep as $u$ (and thus as $v$), and therefore $u$ would not discover any of its neighbors that are marked - and thus $u$ and $v$ would both join $M$. The probability that this event happens is maximized if exactly one pair of neighbors $v$ and $u$ marked themselves and they have no further neighbors that marked themselves. Now the probability that both $v$ and $u$ choose the same subset of half of the time slots of the second interval that has $I$ time slots in total is $1/{\choose{I/2}{I}\leq 1/(I/2)^{I/2}}$. Due to the choice of $I=2^{-2000(\log \degr + \log (2/\eps))}$, we conclude that this is smaller than $2^{-1000(\log \degr + \log (2/\eps))} = \eps\frac{2^{-1000}}{\degr}$. Combining this with the termination probability of the first paragraph of the proof yields that a node terminates and (locally) correctly decides whether it belongs to the MIS with probability $(1-\eps/2)\cdot (1-\eps\frac{2^{-1000}}{\degr})\geq 1-\eps$.
\end{proof}

\iffull
Finally we are ready to prove Theorem~\ref{thm:local-restate} and Theorem~\ref{thm:local-correctness}.
\fi
\begin{proof}(of Theorem~\ref{thm:local-restate2} and Theorem~\ref{thm:local-correctness}). 
We use Theorem~\ref{thm:local-restate} and multiply the number of intervals $R$ by the number $I$ of time steps in each of the two intervals and add the final time slot that completes the emulation. We obtain that the total runtime is
$R\cdot (2I+1) 
O((\log \degr + \log (1/\eps)) \cdot \log(1/\eps)).
$
.
Correctness follows from Theorem~\ref{thm:local-restate} as well.
\end{proof}

\section{Lower Bound on Translating the Schneider and Wattenhofer Algorithm for Bounded Growth Graphs}\label{sec:alg-BGG}
Schneider and Wattenhofer presented an algorithm running in time $O(\log^* N)$ on a restricted class of graphs, see Definition~\ref{def:BGG}, that captures a variety of wireless network topologies. Here, $[1,N]$ indicates the ID space of the $n$ nodes in the network. Theorem~\ref{thm:local-BGG} claims that this algorithm cannot be translated to the \beep model without losing its efficiency  in the following sense: direct translation of the algorithm fails. Major new techniques in the \beep and \local model would need to be developed and the algorithm be modified correspondingly.

\begin{definition}[Bounded growth graphs]\label{def:BGG}
A graph $G=(V,E)$ is \emph{(polynomial) growth-bounded} if there is a polynomial bounding function $f(r)$ such that for each node $v\in V$, the number of nodes in the $r$-neighborhood $N_{G,r}(v)$ of $v$ in $G$ that are in any independent set of $G$ is at most $f(r)$ for all $r\geq 0$. 
\end{definition}

\begin{theorem}\label{thm:local-BGG}
A straightforward implementation of the MIS algorithm of~\cite{DBLP:conf/podc/SchneiderW08} takes $\Omega(\Delta)$ time slots in the \beep model.
\end{theorem}

\ifshort
In the full version of this paper, see Section~\ref{sec:alg-BGG} in the Appendix, we prove Theorem~\ref{thm:local-BGG} and
\fi
\iffull
We 
\fi
provide some intuition behind Theorem~\ref{thm:local-BGG} by reviewing how the algorithm of Schneider and Wattenhofer~\cite{DBLP:conf/podc/SchneiderW08} works and argue that it cannot be translated into the \beep model without losing a $\Delta$ factor, such that its efficiency can not be translated to the \beep model without major modifications. To do so, we construct a network in the plane, in such a way, that the minimum value stored in the neighbors of a node $v$ is different for most nodes $v$ and argue that identifying this value for each node cannot be done fast in the \beep model. This is a key ingredient of the Algorithm by Schneider and Wattenhofer, which therefore does not run fast in the \beep model without major new insights.  

\iffull
Review of algorithm and result of~\cite{DBLP:conf/podc/SchneiderW08}: In the \local model, when applied on Bounded Growth Graphs, the algorithm runs in deterministic time $O(\log^* N)$, where nodes have IDs in the range of $[N]$, which is likely $[$poly $n]$. Recall that in the algorithm of \cite{DBLP:conf/podc/SchneiderW08}, each node is in one of 5 states at any time and stores a value that changes over time. A node $v$ changes its state and value based on the minimum value of $v$'s neighbors. The state change involves a bit by bit comparison between the node's current value and the minimum value around it. Therefore this minimum value needs to be known precisely and cannot be approximated.  

Already Schneider and Wattenhofer~\cite{DBLP:conf/podc/SchneiderW08} point out that the \local model assumes perfect transmission of all messages in each round and therefore their algorithm is less appropriate for wireless networks. We show that this algorithm can not be emulated in the \beep model using less than $\Omega(\Delta)$ slots.

\begin{proof}(Proof sketch of Theorem~\ref{thm:local-BGG}) Consider a graph derived as follows: $\Delta$ nodes $u_1,\dots,u_\Delta$ located equidistantly on a line in ascending order with distance $1$ to each. Let the transmission range be $\Delta/2$ and assume nodes are connected to each other when they are within transmission range. Assume each node $u_i$ has value $2(i-1)+1$ or $2i$, which is determined by an adversary. 

In this graph, the minimum value in the neighborhood of node $u_{\Delta/2+i}$ is $2(i-1)+1$ or $2i$. Node $u_{\Delta/2+i}$ can only determine this value by communicating with node $u_{i}$. This is true for all nodes $u_{\Delta/2+1},\dots,u_\Delta$ and only one pair of nodes can communicate at the same time, such that at least $\Delta/2$ value need to be exchanged. This takes at least $\Omega(\Delta)$ slots.     
\end{proof}
\fi

\section{Discussion and Implications for the Abstract MAC Layer}\label{sec:MAC}
\ifshort
In the full version of the paper, see Appendix, Section~\ref{FULL:sec:MAC}, we
\fi
\iffull
We
\fi
 describe a close connection between the \beep model and abstract MAC layers (a.k.a Local Broadcast Layers) that were introduced by Kuhn et al.~\cite{DBLP:journals/dc/KuhnLN11} and recently got increased attention, e.g., in~\cite{cornejo2009neighbor,cornejo2014reliable,unreliableTechreport,halldorsson2015local-arxiv,halldorsson2015local_podc,DBLP:journals/adhoc/KhabbazianKKL14,DBLP:journals/dc/KuhnLN11,DBLP:conf/podc/Newport14}. We  show how our MIS algorithm can be translated to this model. 
Abstract MAC layers were proposed as a model that provides an alternative approach to the various graph-based models with the goal of abstracting away low level issues with message contention. In this model one can express guarantees  
for local broadcast while hiding the complexities of managing message  
contention. These guarantees include message delivery latency bounds:  
an \emph{acknowledgment bound} $f_{ack}$ on the time for a sender's message to  
be received by \textit{all} neighbors, and a \emph{progress bound} $\fprog$ on the time  
for a receiver to receive \textit{some} message when at least one neighbor is  
sending.

\iffull
Of particular interest with respect to the \beep model is the progress bound. More formally, the progress bound guarantee is as follows: fix some $(u,v)\in E$ and interval of length $\fprog$ throughout which node $u$ is broadcasting a message $m$; during this interval node $v$ must receive some message (though not necessarily $m$, but a message that some location is currently working on).
We consider an enhanced definition of the abstract MAC layer~\cite{DBLP:journals/dc/KuhnLN11}, which provides nodes an abort interface that allows them to abort a broadcast in progress. This is useful, as we can stop a broadcast after time $\fprog$ and know that each node that should receive a message has indeed received one.

We now provide a high-level idea of how to translate our result from the \beep model to the abstract MAC layer. We emulate each slot of the \beep model using $\fprog$ time in the abstract MAC layer. For each round $t$ and for any node $u$ that wants to send a beep in slot $t$, we inject $bcst_v($``beep''$)$ into the MAC layer interface at time $(t-1)\cdot\fprog+1$ and an $abort_v$ command at time $t\cdot\fprog$ to stop the broadcast. Based on the definition of progress, each node $v$ that has a neighbor that sends a message in slot $t$ of the \beep algorithm, received a message ``beep'' in the abstract MAC layer algorithm at time $t\fprog$ at the latest. Using Theorems~\ref{thm:local-correctness} and~\ref{thm:local-restate2} we conclude:
\fi
\begin{theorem}\label{thm:macMIS}
Given an abstract MAC layer that supports aborts and executes the algorithm described above, when a node $v$ terminates, it has made its (locally correct) decision whether it is in the MIS or not, and the probability that node $v$ terminates within the first $O((\log \degr + \log (1/\eps)) \cdot \log(1/\eps)\cdot \fprog)$ slots is at least $1-\eps$. This holds even if the outcome of the coin tosses outside $N^{+}_{2}(v)$ are determined adversarially.
\end{theorem}
\iffull
Thus the cost of our MIS algorithm over abstract MAC depends on the progress 
bound only, not the acknowledgment bound.  Given that, for radio networks at 
least, acknowledgment bounds are much bigger than progress bounds, this produces 
an efficient MIS algorithm for the radio network model.
\fi

\vspace*{-0.2cm}
\addcontentsline{toc}{section}{References} 

{\small
\bibliographystyle{plainurl}
\bibliography{references}
}
\begin{center}
\textbf{APPENDIX}
\end{center}

\section{Basic Chernoff Bounds Used in the Proofs}
\begin{lemma}[Chernoff Bounds~\cite{lecture}]\label{lem:chernoff}
Let $X = \sum_{i=1}^n X_i$, where $X_i = 1$ with probability $p_i$ and $X_i = 0$ with
probability $1-p_i$, and all $X_i$ are independent. Let $\mu = \mathbb{E}[X] = \sum_{i=1}^n p_i$. Then
\begin{enumerate}
\item 
$Pr\left(X\geq (1+\delta)\mathbb{E}[X]\right)\leq e^{-\frac{\delta^2}{2+\delta}\mathbb{E}[X]}$, for all $\delta>0$, and\\
\item 
$Pr\left(X\leq (1-\delta)\mathbb{E}[X]\right)\leq e^{-\frac{\delta^2}{2}\mathbb{E}[X]}$, for all $0<\delta<1$.
\end{enumerate}
\end{lemma}

\end{document}